# Propagation of ion-acoustic solitary waves in a relativistic electron-positron-ion plasma

E. Saberian, A. Esfandyari-Kalejahi and M. Akbari-Moghanjoughi
*Azarbaijan University of Tarbiat Moallem, Faculty of Science, Department of Physics, 51745-406, Tabriz, Iran*
(Dated: November 23, 2010)

E-mail: s.esaberian@azaruniv.edu

**Abstract**

Propagation of large amplitude ion-acoustic solitary waves (IASWs) in a fully relativistic plasma consisting of cold ions and ultrarelativistic hot electrons and positrons is investigated using the Sagdeev's pseudopotential method in a relativistic hydrodynamics model. Effects of streaming speed of plasma fluid, thermal energy, positron density and positron temperature on large amplitude IASWs are studied by analysis of the pseudopotential structure. It is found that in regions that the streaming speed of plasma fluid is larger than that of solitary wave, by increasing the streaming speed of plasma fluid the depth and width of potential well increases and resulting in narrower solitons with larger amplitude. This behavior is opposite for the case where the streaming speed of plasma fluid is smaller than that of solitary wave. On the other hand, increase of the thermal energy results in wider solitons with smaller amplitude, because the depth and width of potential well decreases in that case. Additionally, the maximum soliton amplitude increases and the width becomes narrower as a result of increase in positron density. It is shown that varying the positron temperature does not have considerable effect on the width and amplitude of IASWs. The existence of stationary soliton-like arbitary amplitude waves is also predicted in fully relativistic electron-positron-ion (EPI) plasmas. Effects of streaming speed of plasma fluid, thermal energy, positron density and positron temperature on these kinds of solitons are the same as that for large amplitude IASWs.

## 1. Introduction

It is widely believed that electron-positron (EP) plasmas form the major constituent of several astrophysical situations such as early Universe [1], active galactic nuclei (AGN) [2], pulsar and neutron star atmospheres [3], solar atmosphere [4], accretion disk [5], and black holes [6]. However, there is a possibility that plasma also contain ions besides the electrons and positrons in these environments. In the laboratory experiments, on the other hands, the positron can be introduced in a Tokamak electron-ion (EI) plasma by injecting bursts of neutral positronium atoms ($e^+e^-$) which are then ionized by plasma. The annihilation time of positron in the plasma is long compared to typical particle confinement time. It has been observed experimentally that even at electron density of $10^{12}$ cm$^{-3}$ and temperature as low as 1 eV, the positron annihilation time is greater than one second [7]. Due to long life-time of positron, several low frequency waves can exist in electron-positron-ion (EPI) plasmas which otherwise can not propagate in EP plasmas. Moreover, the response of the EPI plasmas to disturbances is drastically modified. For many years several authors have investigated, theoretically, the nonlinear wave propagation in EPI plasmas to explain some aspects of dynamic behavior of both astrophysical and laboratory plasmas [8-13].

In many space observations and also laboratory experiments in connection with ultra-intense laser pulses interacting with matter, it has been observed that plasma particles attain relativistic speeds [14,15]. It is obvious that in these situations, high-speed streaming ions, electrons and positrons cause the excitation of various kinds of nonlinear structures such as shock and solitary waves. Thus we must take the relativistic motion of plasma particles into consideration in the study of nonlinear plasma waves. Because of encounter with high-speed plasmas both in astrophysical and cosmological applications [16-18] and in laser-plasma interaction research [19-21], many studies in connection with nonlinear phenomena in plasmas have been devoted to the study of relativistic plasmas, such as the study of properties of ion-acoustic solitary waves (IASWs) [22-25] and double layers [26,27] in a relativistic plasma.

Much of earlier works on the study of nonlinear wave motion in relativistic plasmas have been limited to considering weakly relativistic cases [28-32]. However, investigation of large amplitude IASWs in a fully relativistic EPI plasma and the role of various parameters that appear in such plasmas has not yet been studied to authors current knowledge.

The purpose of this paper is to investigate the existence of large amplitude IASWs in a fully relativistic plasma composed of cold ions and ultrarelativistic hot electrons and positrons and to derive the pseudopotential structure for these waves in order to show the dependency of ion-acoustic (IA) waves on number of parameters such as streaming speed of plasma fluid, thermal energy, positron density and positron temperature. The pseudopotential and existence conditions of large amplitude IASWs will be confirmed by considering the ratio of the electron to ion density, the ratio of the positron to ion density, the ratio of the positron to electron temperature and the ratio between the electron thermal energy and ion rest energy in an EPI plasma.

This paper is organized as follows. In sec. II, we present the hydrodynamic model for fully relativistic plasmas and derive the energy equation and pseudopotential in a general form. In sec. III, we derive the appropriate pseudopotential for a relativistic EPI plasma. In sec. IV, we impose conditions and find the least criteria for the existence of large amplitude IASWs on the basis of energy equation. Section V is devoted to discussion and finally, sec. VI contains conclusions of results.

## 2. Hydrodanamic model for relativistic plasmas and formulation

A hydrodynamic model for relativistic plasmas has been established by Lontano *et al.* [33] in 2001 in which they investigate the intraction between arbitary amplitude electromagnetic (EM) fields and hot plasmas. This model has been applied by Lee and Choi [34] in 2007 in order to investigation the large amplitude IASWs in a two-components EI plasma. We apply this model in which the dynamics of the nonlinear waves is governed by a set of equations that is applicable for any arbitary high-speed plasma fluid.

A relativistic multi-component plasma moving in one dimension, say $x$, under the influence of electrostatic (ES) field, described by following set of equations derived from conservation of the particle number and of the energy and momentum in fluid dynamics [33-35]:

$$\frac{\partial}{\partial t}(\gamma n) + \frac{\partial}{\partial x}(\gamma n u) = 0, \tag{1}$$

$$\frac{1}{c^2}\left(\frac{\partial h \gamma u}{\partial t} + u \frac{\partial h \gamma u}{\partial x}\right) + \frac{1}{\gamma n}\frac{\partial p}{\partial x} = -q\frac{\partial \phi}{\partial x}, \tag{2}$$

$$\frac{\partial}{\partial t}\left(\frac{p}{n^\alpha}\right) + u \frac{\partial}{\partial x}\left(\frac{p}{n^\alpha}\right) = 0, \tag{3}$$

and

$$\frac{d^2\phi}{dx^2} = -4\pi \sum_a q_a \gamma_a n_a . \tag{4}$$

Here, $n$ is the particle number density in a reference frame in which the fluid element is at rest at a given instance of time and at a given spatial position, where $u$ is the streaming speed of plasma fluid, $p$ is the pressure and, $\gamma$ is the relativistic factor

$$\gamma = 1/\sqrt{1 - u^2/c^2}, \tag{5}$$

with the enthalpy, h, per fluid particle [34]

$$h = mc^2 + \frac{\alpha}{\alpha - 1} \frac{p}{n}, \tag{6}$$

where, $\alpha$ is the adiabatic index, which is equal to $4/3$ for a hot relativistic gas and $5/3$ for a cold nonrelativistic gas, $q$ is the (invariant) particle charge and $m$ is the rest mass and $\phi$ is the ES potential. The subscript $a$ in Eq. (4) denotes the species of plasma. Eqs. (1)-(3), (5) and (6) apply to each specie of plasma (electrons, positrons and ions), but they are written without the subscript.

Interested reader may refer to appendix A given in Ref. [34] to obtain the equations (1)-(4) for more details. These equations are more accurate for describing fully relativistic fluid model than ones employed by e.g. Nejoh [26] and El-Labany et al. [25]. For instance, the relativistic factor $\gamma$ do not appear in the continuity and Poisson equations given in Refs. [25] and [26]. In addition we note that the pressure equation (3) has been written in conservation law form which is simpler than the pressure equation given in Refs. [25] and [26]. Also, we note that the relativistic effect appears in Eq. (3) by the adiabatic factor $\alpha$.

With these relativistic equations, we use Sagdeev's pseudopotential method [36] in order to investigate the possibility of propagation of large amplitude IASWs.

Unlike of classical approach we use the Lorenz transformation and consider waves in a frame $(\xi, \tau)$ moving along the $x$ axis with (relativistic) velocity $V$ relative to the lab frame. Using Lorenz transformation

$$\xi = \frac{x - Vt}{\sqrt{1 - V^2/c^2}}$$

and

$$\tau = \frac{t - \frac{V}{c^2} x}{\sqrt{1 - V^2/c^2}},$$

in the wave frame, for a stationary wave that is $\partial/\partial \tau \equiv 0$, from Eqs. (1)-(4) we obtain the following relations

$$\frac{\partial}{\partial \xi}[\gamma n(\beta - v)] = 0, \tag{7}$$

$$-v \frac{\partial}{\partial \xi} h\gamma\beta + \beta \frac{\partial}{\partial \xi} h\gamma\beta + \frac{1}{\gamma n} \frac{\partial p}{\partial \xi} = -q \frac{\partial \phi}{\partial \xi}, \tag{8}$$

$$(\beta - v) \frac{\partial}{\partial \xi}\left(\frac{p}{n^\alpha}\right) = 0, \tag{9}$$

and

$$\frac{d^2\phi}{d^2\xi} = -4\pi(1 - v^2)\sum_a q_a \gamma_a n_a, \tag{10}$$

where the dimensionless parameters $\beta$ and $v$ are employed to denote $u/c$ and $V/c$, respectively. For $\beta - v \neq 0$, from Eq. (9) the adiabatic relation is obtained as

$$\frac{p}{n^\alpha} = const. \tag{11}$$

Multiplying Eq. (10) by $d\phi/d\xi$ and recognizing that for localized perturbations (solitons), $\phi$ as well as $d\phi/d\xi$ vanish at $|\xi| \to \infty$, we obtain the energy equation

$$\frac{1}{2}\left(\frac{d\phi}{d\xi}\right)^2 + \Psi = 0, \tag{12}$$

where pseudopotential reads as

$$\Psi = 4\pi(1-v^2)\sum_a q_a \int \gamma_a n_a d\phi. \tag{13}$$

We can use Eqs. (6)-(8) and (11) for each plasma specie with appropriate assumptions to express $\gamma$ as well as $n$ in terms of $\phi$ and then solve the integration in Eq. (13).

## 3. Derivation of pseudopotentian for relativistic electron-positron-ion plasma

Considering a three-components plasma composed of ions, electrons and positrons, the pseudopotential can be written as

$$\Psi = \Psi_i + \Psi_e + \Psi_p. \tag{14}$$

The subscripts $i$, $e$ and $p$ denote ions, electrons and positrons, respectively. With the usual assumption of cold ions, that in relativistic systems means $k_B T_i \ll m_i c^2$, the enthalpy given in Eq. (6) for ions can be approximated as

$$h_i \simeq m_i c^2. \tag{15}$$

Here, $k_B$ is the Boltzmann constant and $T_i$ and $m_i$ are the temperature and rest mass of the ions. We also require that $\phi \to 0$, $\beta \to \beta_0$ and $n \to n_0$ as $|\xi| \to \infty$, for each plasma specie. Integrating Eqs. (7) and (8) with respect to $\xi$ and using Eq. (15), according to Eq. (13) we can calculate the pseudopotential $\Psi_i$ for ions as

$$\Psi_i = 4\pi n_{i0} \gamma_{i0} m_{i0} c^2 (v - \beta_{i0}) \times \left\{ -\frac{e\phi}{m_i c^2} v \pm \left[ \sqrt{\left(\gamma_{i0}(1-\beta_{i0}v) - \frac{e\phi}{m_i c^2}\right)^2 + v^2 - 1} - \gamma_{i0}|\beta_{i0} - v| \right] \right\} \tag{16}$$

With the assumption of ultrarelativistic hot electrons that means $k_B T_e \gg m_e c^2$, calculation of pseudopotential for electrons becomes quite subtle. Here, $T_e$ and $m_e$ are the temperature and rest mass of the electrons. This logical assumption means that massless electrons, in comparison with heavy ions, possess high thermal energy, so that we can neglect the rest energy for electrons and then the enthalpy given in Eq. (6) for electrons can be approximated as

$$h_e \simeq \frac{\alpha}{\alpha-1} \frac{p_e}{n_e}. \tag{17}$$

Integration of Eq. (8) with respect to $\xi$ and using Eq. (11) and the relation $p_{e0} = n_{e0} k_B T_e$ yields

$$n_e = n_{e0} \left(1 + \frac{\alpha}{\alpha-1} \frac{1}{\gamma_{e0}(1-\beta_{e0}v)} \frac{e\phi}{k_B T_e}\right)^{\alpha/\alpha-1}. \tag{18}$$

We mention that hot electrons in plasma move with ultrarelativistic velocities so that perturbations in their velocities in comparison with the ones at infinity are negligible and therefore $\beta_e \simeq \beta_{e0}$ as well as $\gamma_e \simeq \gamma_{e0}$ are held with good approximation.

Therefore, by substituting Eq. (18) into Eq. (13), we obtain the pseudopotential $\Psi_e$ for electrons as

$$\Psi_e = 4\pi n_{e0}\gamma_{e0}^2(1-v^2)(1-\beta_{e0}v)k_BT_e \times \left[1-\left(1+\frac{\alpha-1}{\alpha\gamma_{e0}(1-\beta_{e0}v)}\frac{e\phi}{k_BT_e}\right)^{\alpha/\alpha-1}\right]. \quad (19)$$

Similarly, for positrons with same mass but opposite charge as electrons, we obtain

$$n_p = n_{p0}\left(1-\frac{\alpha}{\alpha-1}\frac{1}{\gamma_{p0}(1-\beta_{p0}v)}\frac{e\phi}{k_BT_p}\right)^{\alpha/\alpha-1}, \quad (20)$$

and the pseudopotential $\Psi_p$ for positrons is calculated as

$$\Psi_p = 4\pi n_{p0}\gamma_{p0}^2(1-v^2)(1-\beta_{p0}v)k_BT_p \times \left[1-\left(1-\frac{\alpha-1}{\alpha\gamma_{p0}(1-\beta_{p0}v)}\frac{e\phi}{k_BT_p}\right)^{\alpha/\alpha-1}\right], \quad (21)$$

where, $T_p$ is the temperature of positrons.

We remind that in Eqs. (19) and (21), $\alpha = 4/3$ which indicates that electrons and positrons are ultrarelativistic hot particles.

## 4. Normalization and existence conditions

Before investigation of existence of large amplitude IASWs, we rewrite the Poisson's equation and pseudopotential in terms of the following normalized parameters

$$\varphi \equiv \frac{e\phi}{m_ic^2}, \quad \psi \equiv \frac{\Psi}{4\pi n_{i0}m_ic^2}, \quad \eta \equiv \frac{\xi}{\lambda_i}, \quad (22)$$

where $\lambda_i$ defined as $\lambda_i = c\sqrt{m_i/4\pi n_{i0}e}$.

The Poisson's equation and pseudopotential take the form

$$\frac{d^2\varphi}{d\eta^2} = -\frac{\partial\psi_\pm}{\partial\varphi} \quad (23)$$

and

$$\psi_\pm = \gamma_{i0}(v-\beta_{i0})\times\left\{-v\varphi \pm \left[\sqrt{(\gamma_{i0}(1-\beta_{i0}v)-\varphi)^2+v^2-1}-\gamma_{i0}|\beta_{i0}-v|\right]\right\}$$

$$+\gamma_{e0}^2(1-v^2)(1-\beta_{e0}v)\varepsilon_e\delta_e \times \left[1-\left(1+\frac{\alpha-1}{\alpha\gamma_{e0}(1-\beta_{e0}v)\varepsilon_e}\varphi\right)^{\alpha/\alpha-1}\right]$$

$$+\gamma_{p0}^2(1-v^2)(1-\beta_{p0}v)\sigma\varepsilon_e\delta_p \times \left[1-\left(1-\frac{\alpha-1}{\alpha\gamma_{p0}(1-\beta_{p0}v)\sigma\varepsilon_e}\varphi\right)^{\alpha/\alpha-1}\right]. \quad (24)$$

Here, the parameters $\delta_e = n_{e0}/n_{i0}$, $\delta_p = n_{p0}/n_{i0}$ and $\sigma = T_p/T_e$ are the ratio of the unperturbed electron and ion densities, the ratio of the unperturbed positron and ion densities and the fractional positron to electron temperature, respectively. We also define the normalized parameter $\varepsilon_e = k_BT_e/m_ic^2$ that is the normalized thermal energy of electrons. It is obvious that $\sigma\varepsilon_e$ is the normalized thermal energy for positrons. Thus in the subsequent discussions we will use $\varepsilon_e$ as a measure of thermal energy of the plasma.

In order for the large amplitude IASWs to exist, the following two conditions must be satisfied:
(i) The pseudopotential must have a local maximum in the point $\varphi = 0$, and the equation $\psi(\varphi) = 0$ should have at least one real solution.

(ii) Nonlinear IASWs exist only when $\psi(\varphi_c) > 0$, where the critical potential $\varphi_c$ is determined by $\varphi_c = \gamma_{i0}(1 - \beta_{i0}v) - \sqrt{1-v^2}$ so that for $\varphi \leq \varphi_c$ the pseudopotential is real.

If these two essential conditions satisfy, in a typical potential well, the soliton starts at the position $\varphi = 0$ with an infinitesimal positive velocity $d\varphi/d\eta$. Then it encounters the potential well in the region $\varphi > 0$ and is reflected at some positive $\varphi = \varphi_m$ for which $\psi(\varphi_m) = 0$, and then return to $\varphi = 0$.

These requirements imply that following conditions must be imposed on pseudopotential:

$$\psi(0) = 0, \tag{25}$$

$$\partial\psi/\partial\varphi|_{\varphi=0} = 0, \tag{26}$$

$$\partial^2\psi/\partial\varphi^2|_{\varphi=0} < 0, \tag{27}$$

and

$$\psi(\varphi_c) > 0. \tag{28}$$

For the pseudopotential given in Eq. (24), the condition $\psi(0) = 0$ is satisfied spontaneously. The condition $\partial\psi/\partial\varphi|_{\varphi=0} = 0$ is in fact the the quasineutrality of plasma at infinity and imply

$$\gamma_{e0}\delta_e = \gamma_i + \gamma_{p0}\delta_p. \tag{29}$$

Considering Eq. (24), we deduce that if $\beta_{i0} > v$, the quasineutrality of plasma at infinity is satisfied for $\psi_+$ branch and if $\beta_{i0} < v$, is satisfied for $\psi_-$ branch of pseudopotential.

We further require that the net electric current due to moving charged particles at infinity has to vanish. Imposing this requirement as a boundary condition and using the current density definition in one dimension, $J = \sum_a q_a \gamma_a n_a u_a$, we obtain

$$\gamma_{e0}\delta_e\beta_{e0} = \gamma_i\beta_{i0} + \gamma_{p0}\delta_p\beta_{p0}. \tag{30}$$

Seeking for conditions in which Eqs. (29) and (30) are compatible, we find that only in the case $\beta_{i0} = \beta_{e0} = \beta_{p0} \equiv \beta_0$, both the quasineutrality of plasma and the absence of net electric current at infinity are satisfied simultaneously. In such a case that equilibrium of plasma at infinity does not disturbed, both Eqs. (29) and (30) reduce to

$$\delta_e = 1 + \delta_p. \tag{31}$$

Now, we are allowed to apply conditions $\partial^2\psi/\partial\varphi^2|_{\varphi=0} < 0$ and $\psi(\varphi_c) > 0$. Applying them to Eq. (24) we derive inequalities

$$(\beta_0 - v)^2(\sigma\delta_e + \delta_p) - \alpha\sigma\varepsilon_e(1 - \beta_0 v)(1 - \beta_0^2) > 0 \tag{32}$$

and

$$(v - \beta_0)(-v\varphi_c \mp \gamma_0|\beta_0 - v|) + \gamma_0(1-v^2)(1 - \beta_0 v)\varepsilon_e\delta_e \times \left[1 - \left(1 + \frac{\alpha - 1}{\alpha\gamma_0(1 - \beta_0 v)\varepsilon_e}\varphi_c\right)^{\alpha/\alpha-1}\right]$$

$$+ \gamma_0(1-v^2)(1 - \beta_0 v)\sigma\varepsilon_e\delta_p \times \left[1 - \left(1 - \frac{\alpha - 1}{\alpha\gamma_0(1 - \beta_0 v)\sigma\varepsilon_e}\varphi_c\right)^{\alpha/\alpha-1}\right] > 0, \tag{33}$$

respectively. As mentioned earlier, we have used $\beta_{i0} = \beta_{e0} = \beta_{p0} \equiv \beta_0$ and $\gamma_0 = 1/\sqrt{1 - \beta_0^2}$. Solving inequalities (32) and (33) simultaneously with applying Eq. (36) leads to regions where solitary wave solutions can exist.

## 5. Results and discussion

Because of complexity of inequalities (32) and (33) to be solved analytically, and dependence of them to several parameters, we analyze them numerically in order to obtain the permitted regions for the existence of IASWs and discuss the pseudopotential structure and its variation with each of parameters. In Figs. 1-3 we have plotted $(\beta_0, \varepsilon_e)$ plane, $(\beta_0, \delta_p)$ plane and $(\beta_0, \sigma)$ plane for $v = 0.3$ and $\alpha = 4/3$ that show the regions where inequalities (32) and (33) are satisfied simultaneously. In each of these planes, two other parameters are held fixed. These figures reveal that the permitted regions are divided into two branches: the left branch correspond to $\psi_-$ and the right branch correspond to $\psi_+$. We find explicitly that in left branch, the streaming speed of plasma fluid is smaller than solitary wave speed ($\beta_0 < v$), and in the right branch, the streaming speed of plasma fluid is greater than solitary wave speed ($\beta_0 > v$). In each graph the solid and dashed curve represent the condition $\partial^2 \psi / \partial \varphi^2 |_{\varphi=0} = 0$ and $\psi(\varphi_c) = 0$, respectively and the confined area between two curves, shaded regions, are the permitted regions where IASWs can exist.
These graphs also indicate the effect of each plasma parameter on formation of potential well and hence on propagation of large amplitude IASWs.

### 5.1. Effect of streaming speed of plasma fluid

It is observed from Figs. 1-3 that for constant values of $\varepsilon_e$, $\delta_p$ and $\sigma$, increasing the streaming speed of plasma causes different results, depending on whether we consider the left or right branch of pseudopotential. In the left branch, $\beta_0 < v$, with increasing the streaming speed of plasma the typical point in each graph approaches to solid curve, therefore magnitude of $\partial^2 \psi / \partial \varphi^2 |_{\varphi=0}$ approaches to small values and therefore we expect the depth and width of potential well to decrease. This result is opposite in the right branch, $\beta_0 > v$, where with increasing the streaming speed of plasma we have larger magnitude of $\partial^2 \psi / \partial \varphi^2 |_{\varphi=0}$ and so we anticipate the depth and width of potential well to increase. We keep in mind that a deeper and wider potential well corresponds to a narrower and larger solitary wave, which is explicitly represented in Ref. 34.
In Figs. 4 and 5 we show the brid's eye view of the pseudopotential $\psi$ by numerical calculation in the case $v = 0.3$, $\alpha = 4/3$, $\delta_p = 0.01$, $\sigma = 0.3$ and $\varepsilon_e = 0.0043$, that exhibit the effect of streaming speed of plasma fluid on left and right branches of $\psi$, respectively. Figs. 6 and 7 also illustrate the dependency of $\psi$ on the electrostatic potential for some values of $\beta_0$, depending on that the fluid velocity is smaller or larger than solitary wave speed. The value $\varepsilon_e = 0.0043$ corresponds to $k_B T_e \simeq 8 m_e c^2$ or temperatures of order $T_e \simeq 4.6 \times 10^{10}~K$, which has been observed in high-power laser experiments [37]. For $\varepsilon_e = 0.0043$ the allowable range of $\beta_0$ is $0.177 < \beta_0 < 0.230$ for $\psi_-$ branch and $0.365 < \beta_0 < 0.411$ for $\psi_+$ branch of pseudopotential.

### 5.2. Effect of thermal energy

We find from Fig. 1 that regardless of streaming speed of plasma relative to solitary wave speed, as the thermal energy increases the typical point $(\beta_0, \varepsilon_e)$ approaches to solid curve and therefore we conclude that the depth and width of potential well decreases. Thus with increasing the thermal energy, wider solitons with smaller maximum amplitude can propagate in plasma. The effect of thermal energy on pseudopotential structure is shown in a brid's eye view of $\psi$ in Fig. 8 when $v = 0.3$, $\alpha = 4/3$, $\delta_p = 0.01$, $\sigma = 0.3$ and $\beta_0 = 0.1$. The allowable range of $\varepsilon_e$ for $\beta_0 = 0.1$ is $0.0108 < \varepsilon_e < 0.0325$. An example of astrophysical phenomena that belongs to this range of thermal energy is jet cores of active galactic nuclei, for which it is estimated that $T_e \simeq 2.8 \times 10^{11} K$ or equivalently $\varepsilon_e = 0.026$ [38].

**5.3. Effect of positron density**

We can investigate the effect of positron density on pseudopotential structure and IASWs from Fig. 2 Similar to discussions in previous subsections, this graph indicates that for a given thermal energy, an increase in positron density causes the potential well to become deeper and wider, because of large values of $\partial^2 \psi / \partial \varphi^2 |_{\varphi=0}$. Fig. 9 shows a brid's eye view of pseudopotential in the case of $v = 0.3$, $\alpha = 4/3$, $\sigma = 0.3$, $\beta_0 = 0.22$ and $\varepsilon_e = 0.0043$ which demonstrates the variation of $\psi$ with the change of positron density. It is concluded that in a relativistic EPI plasma, amplitude of IA solitary wave becomes larger and the pulse becomes narrower as a result of an increase in positron density. This result is different from corresponding result in nonrelativistic case in which the presence of the positron component reduces the IA soliton amplitude [8].

**5.4. Effect of positron temperature**

From Fig. 3 we deduce that in allowed regions in $(\beta_0, \sigma)$ plane, variation of positron temperature doesn't alter the value of $\partial^2 \psi / \partial \varphi^2 |_{\varphi=0}$ considerably, therefore the variations in potential well and IASWs in terms of changes in positron temperature are negligible. This result is seen explicitly in a brid's eye view of pseudopotential in Fig. 10 where $v = 0.3$, $\alpha = 4/3$, $\delta_p = 0.01$, $\beta_0 = 0.22$ and $\varepsilon_e = 0.0043$. Such a behavior is different from what is observed in a nonrelativistic EPI plasma, in which the large amplitude IASWs sensitively depend on positron temperature [39].

**5.5. Stationary soliton-like structures**

An interesting feature of relativistic plasmas is that they contain special types of solitons that are immobile soliton-like patterns. The existence of such stationary localized structures with arbitary amplitude was predicted by Lontano *et al*. [33]. in a relativistic hot plasma composed of EP pairs in EM fields and thereafter by Lee and Choi [34] in a relativistic EI plasma in ES field. These kinds of solitons are predicted to exist in overdense plasmas [33].

Here, we solve inequalities (32) and (33) simultaneously in the case of $v = 0$ and obtain the existence regions for arbitary amplitude stationary IA solitons in a relativistic EPI plasma. Figs. 11-13 show the existence regions for these solitons which we can analyze in the same way as we proceeded in previous subsections. It is found that increasing the streaming speed of plasma causes a deeper and wider potential well and thus a narrower and larger soliton. Fig. 11 indicates that as the thermal energy increases the depth and width of potential well decreases. It is also found from Fig. 12 that with increasing the positron density the potential well becomes deeper and wider. Furthermore from Fig. 13, it is observed that variation in positron temperature does not significantly affect the potential well and soliton-like structures. A typical potential well for these solitons is shown in Fig. 14 where $\alpha = 4/3$, $\delta_p = 0.01$, $\sigma = 0.3$, $\beta_0 = 0.27$ and $\varepsilon_e = 0.026$.

**6. Conclusions**

In this paper we have obtained the criteria of the existence for large amplitude ion-acoustic solitary waves (IASWs) in a fully relativistic plasma composed of cold ions and hot electrons and positrons. A relativistic hydrodynamic model, limited to one dimension, is applied in order to derive the pseudopotential and energy equation with the purpose of analyzing the pseudopotential structure. To extract more information as how various parameters influence the depth and width of potential well and so how the pulse width and maximum amplitude of solitons vary, we performed the numerical analysis and displayed the typical results in the form of graphs. The results are summarized as follows:

(1) The nonlinear IASWs can exist in the fully relativistic electron-positron-ion (EPI) plasma. It is observed that in regions that the streaming speed of plasma fluid is larger than the solitary wave speed, the depth and width of potential well increases with increasing the streaming speed of plasma and thus we have a larger and narrower solitary wave. This behavior is opposite in the case where the streaming speed of plasma fluid is smaller than the solitary wave speed.

(2) It is found that regardless the value of the streaming speed of plasma fluid relative to solitary wave speed, the depth and width of potential well decreases with increasing the thermal energy and therefore the pulse width and maximum amplitude of solitons decrease.

(3) It is revealed that increasing the positron density enhances the depth and width of potential well and hence narrower solitons with larger amplitude can propagate. This result is different from corresponding result in nonrelativistic case.

(4) It is also shown that positron temperature does not have considerable effect on formation of potential well and therefore varying the positron temperature does not cause a prominent change in width and amplitude of solitons. This behavior differs from what occurs in nonrelativistic EPI plasmas.

(5) The last section of this paper predicts the existence of stationary soliton-like structures of arbitary amplitude in a relativistic EPI plasma. Effects of streaming speed of plasma fluid, thermal energy, positron density and positron temperature on these kinds of solitons are the same as that for IASWs.

We believe that the present study may provide a deeper insight in understanding the nonlinear propagation of electrostatic perturbations which frequently occur in astrophysical as well as experimental plasma environments.

## Acknowledgments

This work has been financially supported by Research Office of Azarbaijan University of Tarbiat Moallem, Tabriz, Iran.

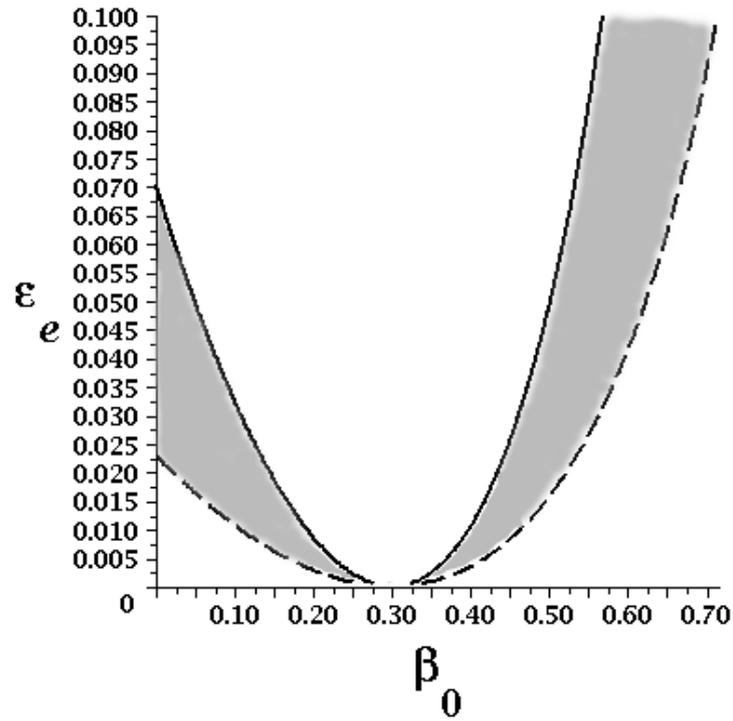

**Fig. 1.** The permitted regions for the existence of ion-acoustic solitary waves in $(\beta_0, \varepsilon_e)$ plane in the case $v = 0.3$, $\alpha = 4/3$, $\delta_p = 0.01$ and $\sigma = 0.3$.

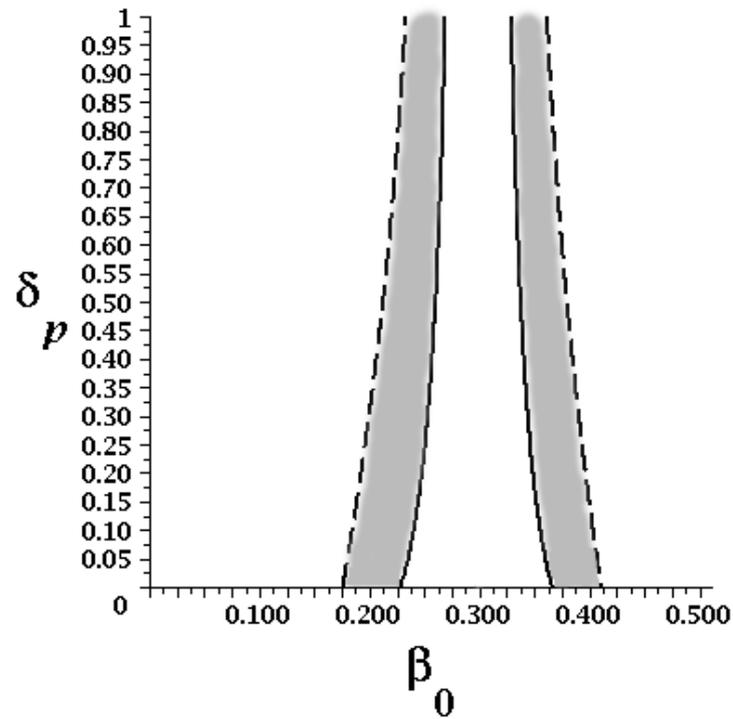

**Fig. 2.** The permitted regions for the existence of ion-acoustic solitary waves in $(\beta_0, \delta_p)$ plane in the case $v = 0.3$, $\alpha = 4/3$, $\varepsilon_e = 0.0043$ and $\sigma = 0.3$.

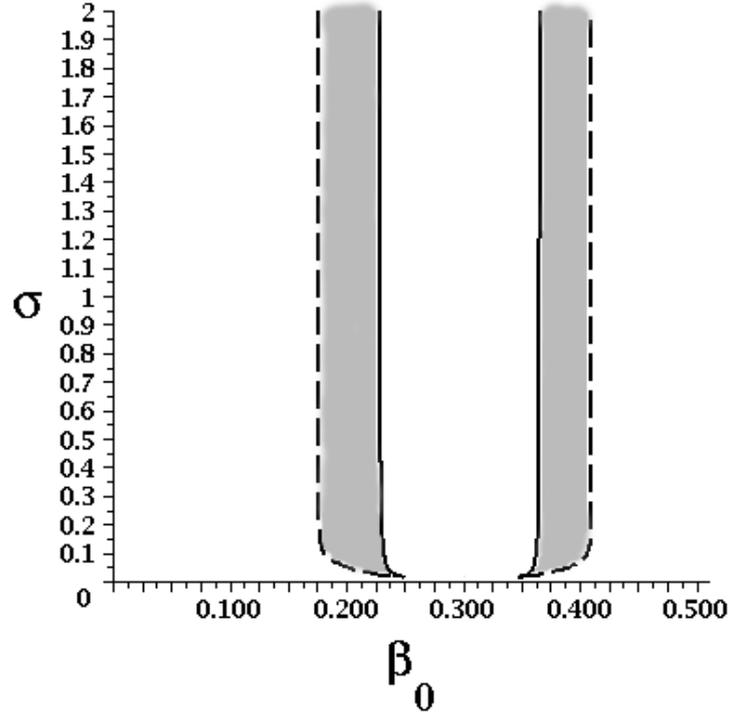

**Fig. 3.** The permitted regions for the existence of ion-acoustic solitary waves in $(\beta_0,\sigma)$ plane in the case $v=0.3$, $\alpha=4/3$, $\varepsilon_e=0.0043$ and $\delta_p=0.01$.

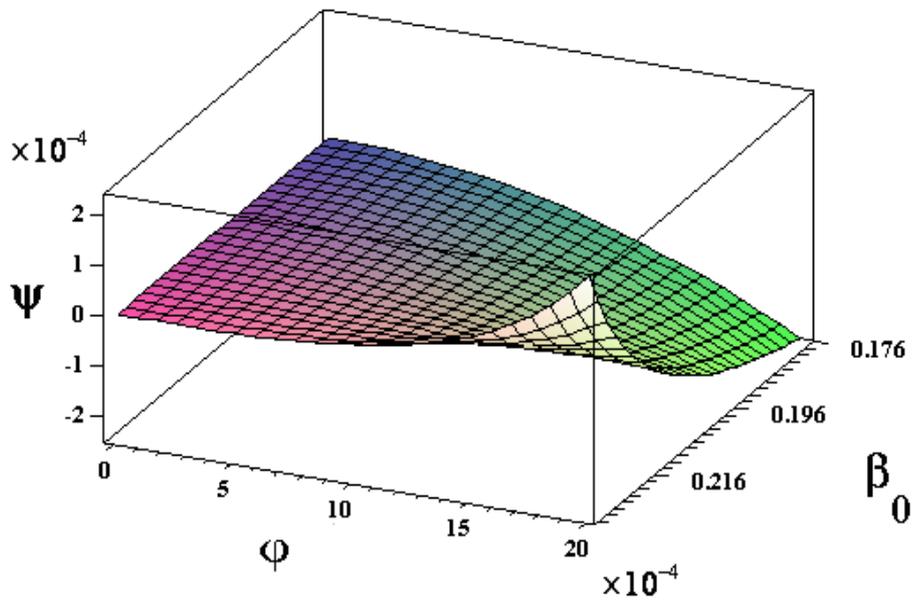

**Fig. 4.** (Color online) Brid's eye view of the pseudopotential for large amplitude ion-acoustic solitary waves that exhibits the effect of streaming speed of plasma on left branch of $\psi$ in the case $v=0.3$, $\alpha=4/3$, $\delta_p=0.01$, $\sigma=0.3$ and $\varepsilon_e=0.0043$.

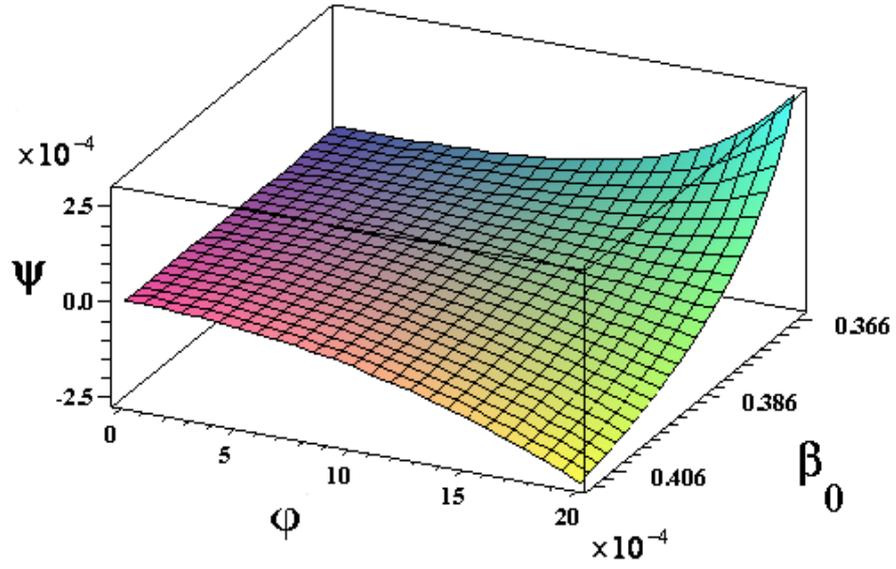

**Fig. 5.** (Color online) Brid's eye view of the pseudopotential for large amplitude ion-acoustic solitary waves that exhibits the effect of streaming speed of plasma on right branch of $\psi$ in the case $v = 0.3$, $\alpha = 4/3$, $\delta_p = 0.01$, $\sigma = 0.3$ and $\varepsilon_e = 0.0043$.

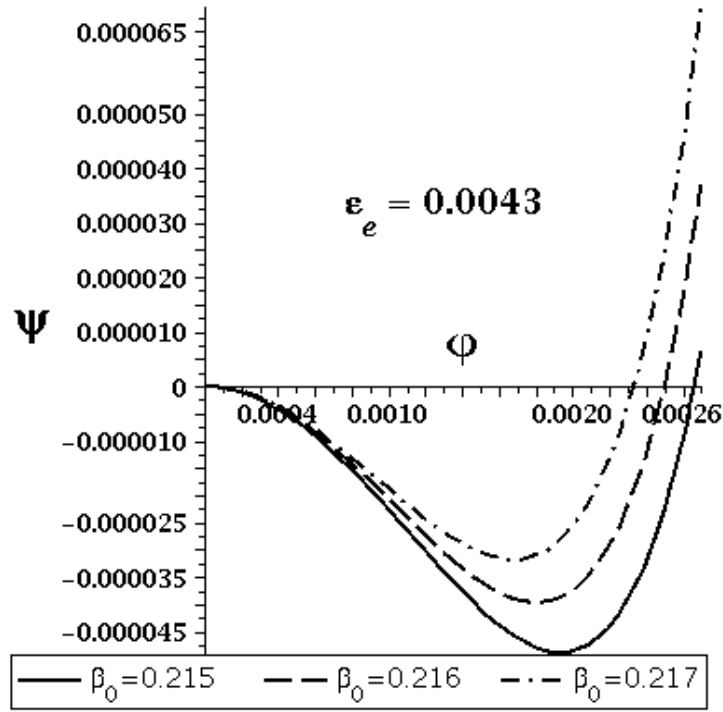

**Fig. 6.** Dependency of $\psi$ on the electrostatic potential for some values of $\beta_0$ in the case that streaming speed of plasma is smaller than solitary wave speed.

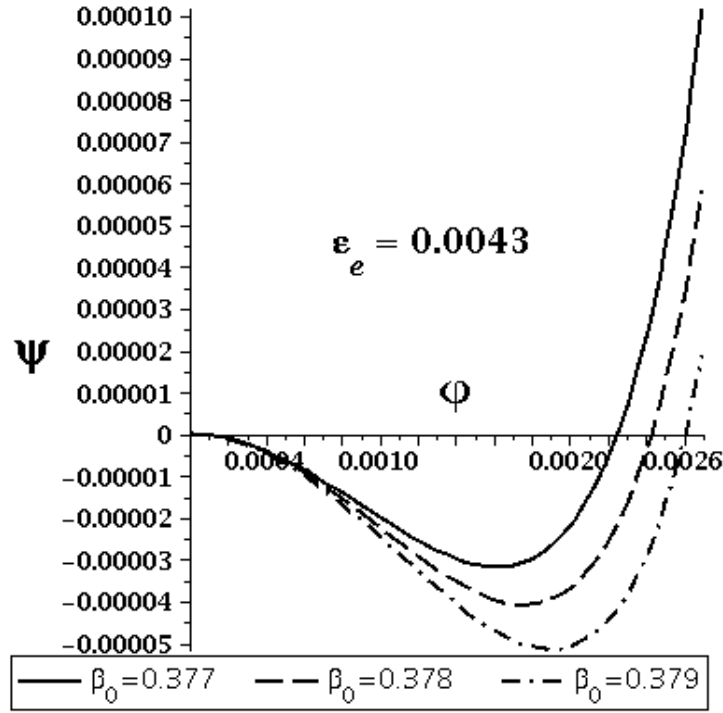

**Fig. 7.** Dependency of $\psi$ on the electrostatic potential for some values of $\beta_0$ in the case that streaming speed of plasma is larger than solitary wave speed.

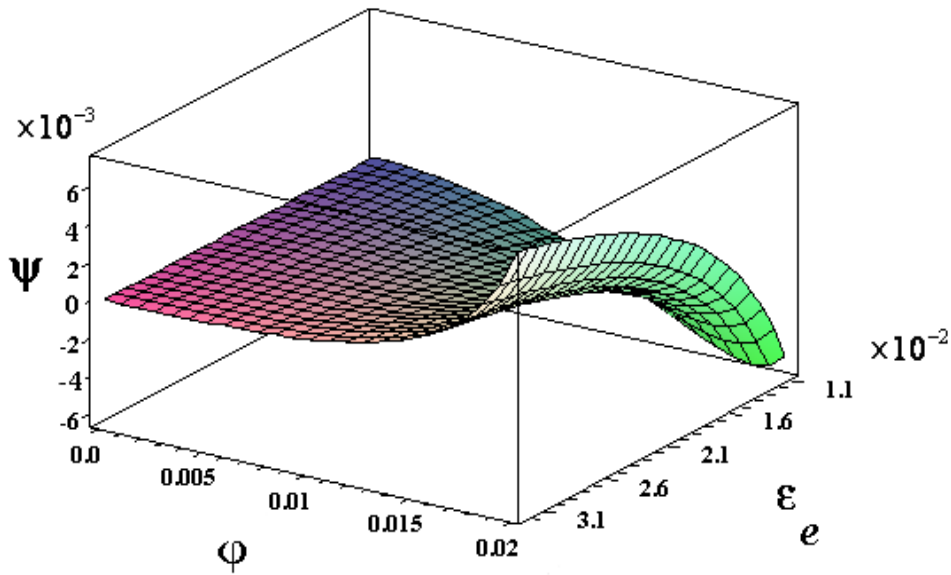

**Fig. 8.** (Color online) Brid's eye view of the pseudopotential for large amplitude ion-acoustic solitary waves that exhibits the effect of thermal energy on potential well in the case $v = 0.3$, $\alpha = 4/3$, $\delta_p = 0.01$, $\sigma = 0.3$ and $\beta_0 = 0.1$.

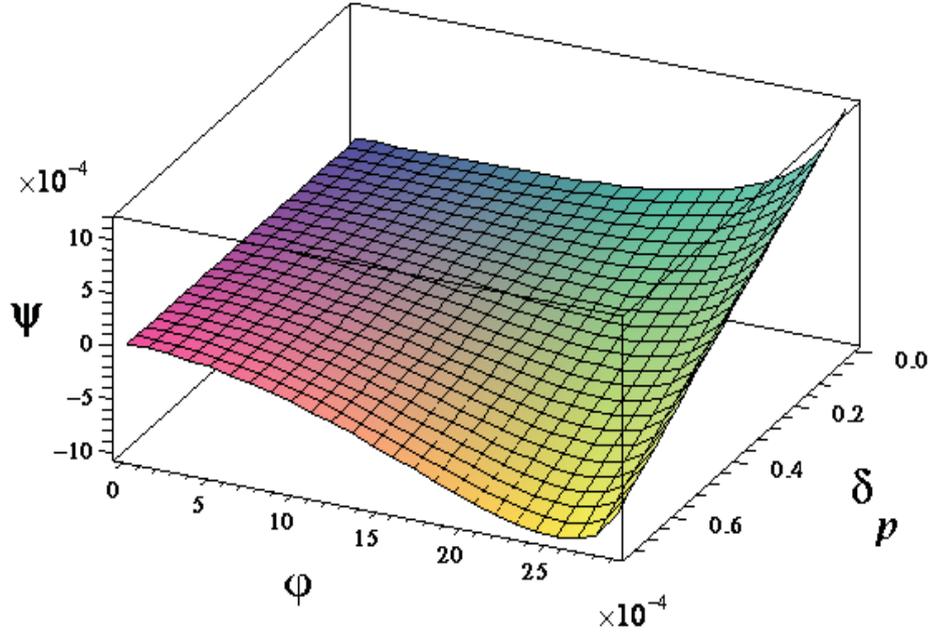

**Fig. 9.** (Color online) Brid's eye view of the pseudopotential for large amplitude ion-acoustic solitary waves that exhibits the effect of positron density on potential well in the case $v = 0.3$, $\alpha = 4/3$, $\sigma = 0.3$, $\beta_0 = 0.22$ and $\varepsilon_e = 0.0043$.

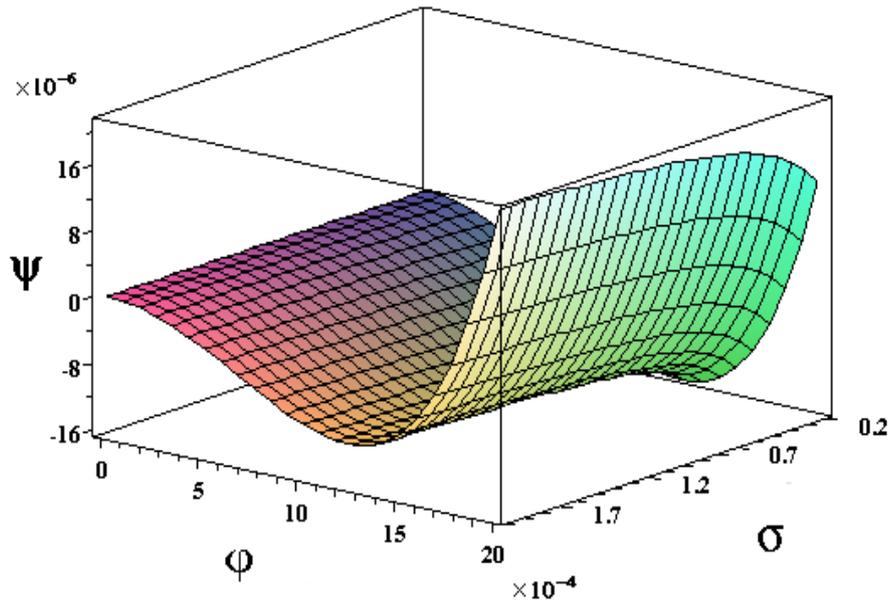

**Fig. 10.** (Color online) Brid's eye view of the pseudopotential for large amplitude ion-acoustic solitary waves that exhibits the effect of positron temperature on potential well in the case $v = 0.3$, $\alpha = 4/3$, $\delta_p = 0.01$, $\beta_0 = 0.22$ and $\varepsilon_e = 0.0043$.

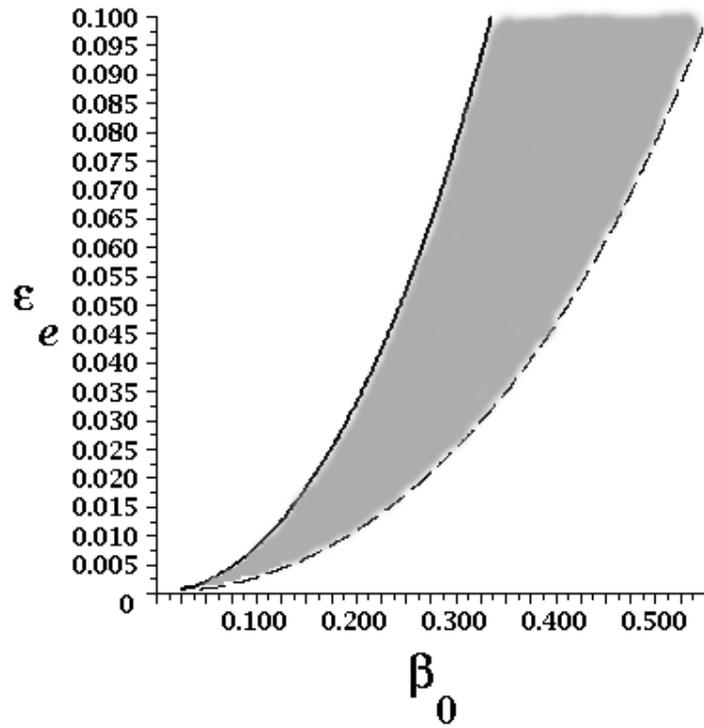

**Fig. 11.** The permitted regions for the existence of stationary soliton-like structures of arbitary amplitude in $(\beta_0, \varepsilon_e)$ plane in the case $\alpha = 4/3$, $\delta_p = 0.01$ and $\sigma = 0.3$.

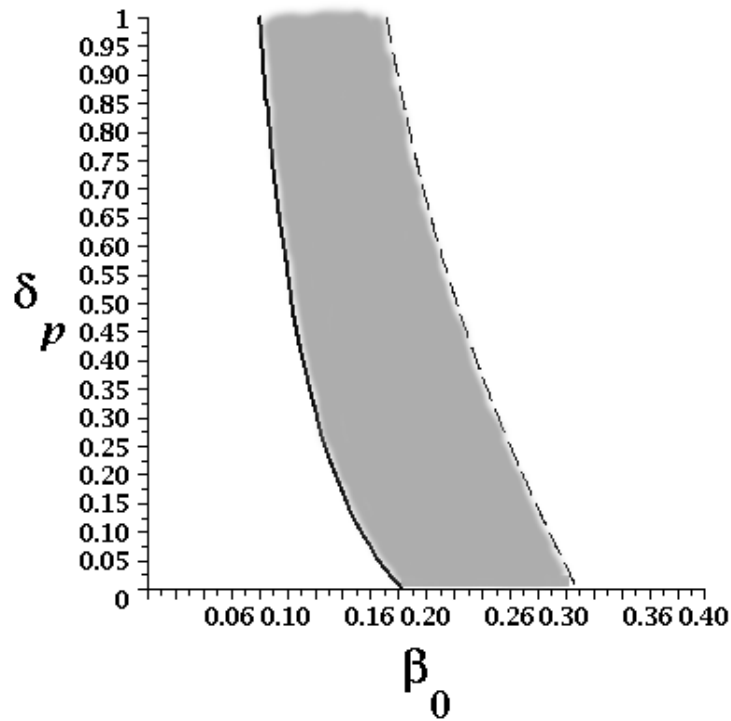

**Fig. 12.** The permitted regions for the existence of stationary soliton-like structures of arbitary amplitude in $(\beta_0, \delta_p)$ plane in the case $\alpha = 4/3$, $\varepsilon_e = 0.026$ and $\sigma = 0.3$.

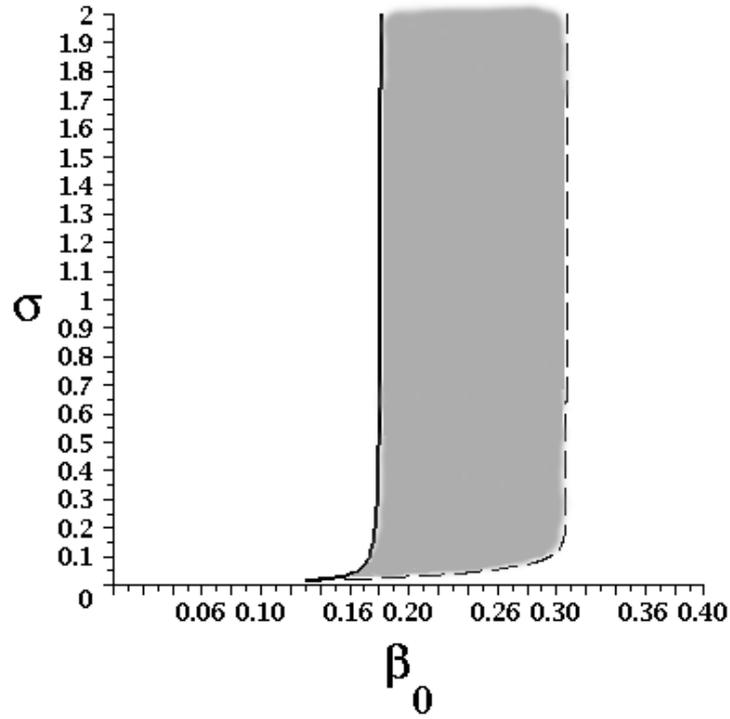

**Fig. 13.** The permitted regions for the existence of stationary soliton-like structures of arbitary amplitude in $(\beta_0, \sigma)$ plane in the case $\alpha = 4/3$, $\varepsilon_e = 0.026$ and $\delta_p = 0.01$.

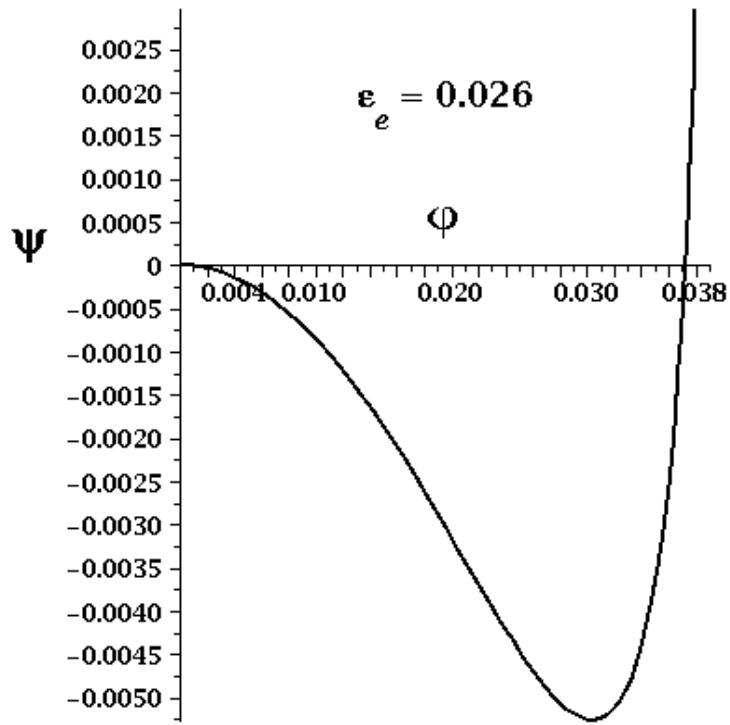

**Fig. 14.** The typical potential well for stationary ion-acoustic solitons in the case $\alpha = 4/3$, $\delta_p = 0.01$, $\sigma = 0.3$, $\beta_0 = 0.27$ and $\varepsilon_e = 0.026$.